# A New Class of Generalized Burr III Distribution for Lifetime Data

[1]Olobatuyi, K. I., [2]Asiribo, O. E., and [3]Dawodu, G. A.
*Federal University of Agriculture, Abeokuta*
*Department of Statistics*
[1]Email: *olobatuyikenny@gmail.com*

**Abstract**

*For the first time, the Generalized Gamma Burr III (GGBIII) is introduced as an important model for problems in several areas such as actuarial sciences, meteorology, economics, finance, environmental studies, reliability, and censored data in survival analysis. A review of some existing gamma families have been presented. It was found that the distributions cannot exhibit complicated shapes such as unimodal and modified unimodal shapes which are very common in medical field. The Generalized Gamma Burr III (GGBIII) distribution which includes the family of Zografos and Balakrishnan as special cases is proposed and studied. It is expressed as the linear combination of Burr III distribution and it has a tractable properties. Some mathematical properties of the new distribution including hazard, survival, reverse hazard rate function, moments, moments generating function, mean and median deviations, distribution of the order statistics are presented. Maximum likelihood estimation technique is used to estimate the model parameters and applications to real datasets in order to illustrate the usefulness of the model are presented. Examples and applications as well as comparisons of the GGBIII to the existing Gamma-G families are given.*

## 1      Introduction

It is known that the Burr III distribution is the third example of solutions of the differential equation defining the Burr system of distribution, (Burr, 1942). This distribution has been used widely in numerous fields of sciences with different parameterizations using other names. For example, it is used as inverse Burr distribution in the actuarial literature III distribution in low-flow frequency analysis where the lower tail of a distribution is of interest. (Klugman et al., 1998) and kappa distribution in the meteorological literature (Mielke, 1973; Mielke and Johnson, 1973). The Burr III distribution has been useful in financial literature, environmental studies, in survival and reliability theory, (Sherrick et al.,1996; Lindsay et al., 1996; Al-Dayian, 1999; Shao, 2000; Hose, 2005; Mokhlis, 2005; Gove *et al.,* 2008). Recently, Shao *et al.,* 2008 proposed the use of the so-called extended Burr.

Burr (1942) introduced a system of distributions which contains the Burr XII (BXII) distribution as the most widely used of these distributions. If a random variable $X$ has the BXII distribution, then $X^{-1}$ has the scaled Burr III (BIII) distribution with cumulative distribution function (cdf) defined (for $x > 0$) by (Antonio et al., 2014)

$$F(x;\beta,\delta,\lambda) = \left(1 + \left(\frac{x}{\lambda}\right)^{-\delta}\right)^{-\beta}, \quad x > 0, \tag{1}$$

where $\beta > 0$, $\delta > 0$ and $\lambda > 0$, are shapes and scale parameters respectively. The probability density function corresponding to (1) is given by

$$f(x; \beta, \delta, \lambda) = \beta \lambda^\delta \delta y^{-\delta-1} \left(1 + \left(\frac{x}{\lambda}\right)^{-\delta}\right)^{-\beta-1}, \qquad (2)$$

The hazard and reverse hazard functions are given by:

$$h(x; \beta, \delta, \lambda) = \frac{\beta \lambda^\delta \delta x^{-\delta-1} \left(1 + \left(\frac{x}{\lambda}\right)^{-\delta}\right)^{-\beta-1}}{1 - \left(1 + \left(\frac{x}{\lambda}\right)^{-\delta}\right)^{-\beta}}, \qquad (3)$$

and

$$\tau(x; \beta, \delta, \lambda) = \beta \lambda^\delta \delta y^{-\delta-1} \left(1 + \left(\frac{x}{\lambda}\right)^{-\delta}\right)^{-1} \qquad (4)$$

The $r^{th}$ raw or non-central moments are:

$$E(Y^r) = \beta^* \lambda^r B\left(\beta^* + \frac{r}{\delta}, 1 - \frac{r}{\delta}\right), \qquad r < \delta \qquad (5)$$

Where $B(a, b) = \frac{\Gamma(a)\,\Gamma(b)}{\Gamma(a+b)}$.

In this paper, we present a new class of Burr-type distribution called the Generalized Gamma-Burr III (GGBIII) distribution and apply the model to some real-life situations.

Motivated by the various applications of Burr III distribution in finance and actuarial sciences, and economics, where distribution plays an important role in size distribution (Antonio *et al.*, 2014). It is customary to develop models that take into consideration not only shape, and scale but also skewness, kurtosis and tail variations. An obvious reason for generalizing a standard distribution is to provide larger flexibility in modeling real data. It is well known in general that a generalized model is more flexible than the ordinary model and it is preferred by many data analysts in analyzing statistical data, (Ojo and Olapade, 2005). The gamma distribution is the most effective model for analyzing skewed data (Marcelino *et al.*, 2011).

In the last few years, several ways of generating new probability distributions from classic ones were developed and discussed. Jones (2004) studied a distribution family that arises naturally from the distribution of order statistics. The beta-generated family proposed by Eugene et al., (2002) has been extensively used by many researchers in generalizing distribution. For any baseline continuous distribution $F(x)$ with survival function $1 - F(x)$ and density $f(x)$, Zografos and Balakrishnan (2009) defined the cumulative distribution function (cdf) and probability density function (pdf) as follows

$$G(x) = \frac{1}{\Gamma(\alpha)} \int_0^{-\log(1-F)} t^{\alpha-1} e^{-t} dt, \quad x \in \mathbf{R}, \quad \alpha > 0 \qquad (6)$$

and

$$g(x) = \frac{1}{\Gamma(\alpha)} \left[-\log(1 - F(x))\right]^{\alpha-1} f(x), \qquad (7)$$

respectively. Also, Ristic and Balakrishnan (2011) proposed an alternative gamma-generator defined by the cdf and pdf given by

$$G(x) = 1 - \frac{1}{\Gamma(\alpha)} \int_0^{-\log F(x)} z^{\alpha-1} e^{-z} dz, \qquad z \in \mathbf{R}, \alpha > 0, \tag{8}$$

and

$$g(x) = \frac{1}{\Gamma(\alpha)} [-\log F(x)]^{\alpha-1} f(x), \tag{9}$$

respectively. Ristic and Balakrishnan (2012) generalized the exponentiated exponential (EE) distribution and further studied by Gupta and Kundu (1999) with cdf $F(x) = (1 - e^{-\beta x})^\alpha$, where $\alpha > 0$ and $\beta > 0$ inserted into equation (8) to obtain and study the gamma exponentiated exponential (GEE) model. Luis *et al.,* (2012) presented the statistical properties of the gamma-exponentiated Weibull distribution.

Broderick et al., (2014) defined the distribution with pdf $g(x)$ and cdf $G(x)$ for any baseline cdf $F(x)$, and $x \in \mathbb{R}$, (for $\alpha > 0$) as follows

$$g(x) = \frac{1}{\Gamma(\alpha)\theta^\alpha} [-\log(1 - F(x))]^{\alpha-1} (1 - F(x))^{(1/\theta)-1} f(x), \tag{10}$$

$$G(x) = \frac{1}{\Gamma(\alpha)\theta^\alpha} \int_0^{-\log(1-F(x))} z^{\alpha-1} e^{-z/\theta} dz, \tag{11}$$

respectively, for $\alpha, \theta > 0$, where $g(x) = \frac{dG(x)}{dx}$, $\Gamma(\alpha) = \int_0^\infty z^{\alpha-1} e^{-z} dz$ denotes the gamma function, and $\gamma(t, \alpha) = \int_0^t z^{\alpha-1} e^{-z} dz$, denotes the incomplete gamma function. Towards the end, they obtained a natural extension for Dagum distribution, which they called the gamma-Dagum (GD) distribution.

A review of some existing gamma families have been presented, Ristic and Balakrishnan (2012) introduced a different type of gamma generalized distribution, Zografos and Balakrishnan (2009) provided a Zografos and Balakrishnan-Dagum (ZB-D) which was modified by Broderick et al. (2014). It was found that these distributions cannot exhibit complicated shapes such as unimodal and modified unimodal shapes which are very common in medical field.

In this paper, we obtain the natural extension of Burr III distribution which we call the Generalized Gamma Burr III distribution.

This paper is organized as follows. In section 2, some basic results, the generalized gamma-Burr III (GGBIII) distribution, series expansion and its sub-models, hazard, survival and reverse hazard functions and the quantile function are presented. The moments and moment generating function, mean and median deviations are given in section 3. Section 4 contains some additional useful results on the distribution of order statistics. In section 5, results on the estimation of the parameters of the GGBIII distribution via the method of maximum likelihood are presented. Applications in different fields are presented in Section 6, followed by concluding remarks in section 7 and references in section 8.

## 2     The Generalized Gamma-G Family

The standard Gamma distribution of different kind is given by Gradshteyn and Ryzhik (2000)

$$\int_0^\infty z^{\omega-1} e^{-z^p/\theta} dz = \frac{\Gamma(\omega/p)\theta^{\omega/p}}{p} \tag{12}$$

Then, we write equation (12) as

$$\frac{1}{\Gamma(\alpha)\theta^\alpha} \int_0^x (z^p)^{\alpha-1} e^{-z^p/\theta} dz \qquad (13)$$

This is the known gamma distribution when $p = 1$

Equation (13) has the parent distribution with the cdf given as follows

$$G(x) = \frac{1}{\Gamma(\alpha)\theta^\alpha} \int_0^{[-\log(1-F(x))]^p} z^{\alpha-1} e^{-z/\theta} dz \qquad (14)$$

Equation (14) is the generalization of Gamma-G family of Broderick et al., (2014) in equation (11) when $p = 1$ and also Zografos and Balakrishnan (2009) in equation (6) when $p = \theta = 1$.

Let $F(x)$ be the cumulative distribution function (CDF) of any random variable $X$. The cdf and pdf of the generalized gamma-G family of distributions when $\theta = 1$ is given by

$$G(x) = \frac{1}{\Gamma(\alpha)} \int_0^{[-\log(1-F(x))]^p} z^{\alpha-1} e^{-z} dz \qquad (15)$$

Differentiating the equation (15) we obtain the pdf as follows

$$g(x) = \frac{1}{\Gamma(\alpha)} \left[[-\log(1-F(x))]^p\right]^{\alpha-1} e^{-[-\log(1-F(x))]^p} \frac{d[-\log(1-F(x))]^p}{dx} \qquad (16)$$

$$g(x) = \frac{1}{\Gamma(\alpha)} \left[[-\log(1-F(x))]^p\right]^{\alpha-1} e^{-[-\log(1-F(x))]^p} p[-\log(1-F(x))]^{p-1} \frac{f(x)}{1-F(x)} \qquad (17)$$

respectively, $\alpha, \beta, \lambda, \delta, p > 0$.

### 2.1 Generalized Gamma Burr III distribution

The new model is proposed by inserting scaled Burr III distribution into equation (15), the cdf $G_{GGBIII}(x)$ of the Generalized Gamma-BIII distribution is obtained as follows:

$$G(x) = \frac{1}{\Gamma(\alpha)} \int_0^{\left[-\log\left(1-\left(1+\left(\frac{x}{\lambda}\right)^{-\delta}\right)^{-\beta}\right)\right]^p} z^{\alpha-1} e^{-z} dz \qquad (18)$$

$$= \frac{\gamma\left(\left[-\log\left(1-\left(1+\left(\frac{x}{\lambda}\right)^{-\delta}\right)^{-\beta}\right)\right]^p, \alpha\right)}{\Gamma(\alpha)} \qquad (19)$$

where $\gamma(x, \alpha) = \int_0^x z^{\alpha-1} e^{-z} dz$ is the incomplete gamma function.

The probability density function is given by

$$g(x) = \frac{p\beta\delta\lambda^{\delta}x^{-\delta-1}}{\Gamma(\alpha)}\left(1+\left(\frac{x}{\lambda}\right)^{-\delta}\right)^{-\beta-1}\left[\left[-\log\left(1-\left(1+\left(\frac{x}{\lambda}\right)^{-\delta}\right)^{-\beta}\right)\right]^{p}\right]^{\alpha-1}$$

$$\times e^{-\left[-\log\left(1-\left(1+\left(\frac{x}{\lambda}\right)^{-\delta}\right)^{-\beta}\right)\right]^{p}}\left[1-\left(1+\left(\frac{x}{\lambda}\right)^{-\delta}\right)^{-\beta}\right]^{-1} \qquad (20)$$

**Shape of GGBIII Distribution**

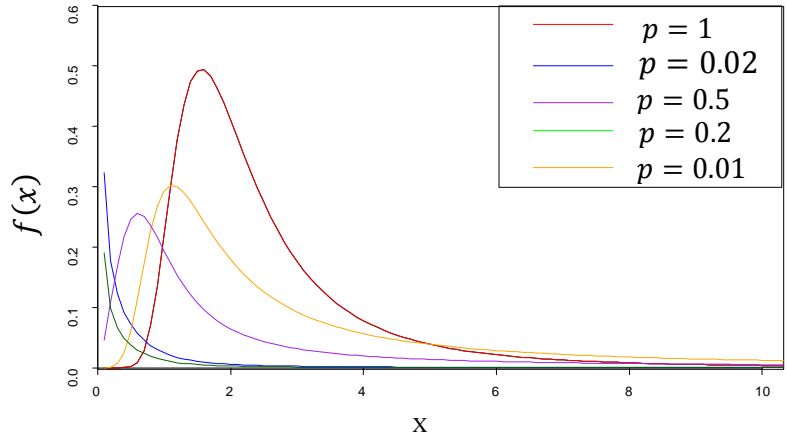

Figure 1: Shape of GGBIII distribution for selected parameters

Plots of the density function of the Generalized Gamma Burr III distribution for selected parameters values are given in figure 1. The plot indicates that the GGBIII distribution can be decreasing or right skewed.

## 2.2 Expansions of density functions

If a random variable $X$ has the GGBIII density, we write $X \sim GGBIII(\alpha, \beta, \delta, \lambda, p)$. Let $u = \left(1+\left(\frac{x}{\lambda}\right)^{-\delta}\right)^{-\beta}$, and then using the series representation, we have

$$\log\frac{1}{(1-u)} = \sum_{k=1}^{\infty}\frac{u^{k}}{k} = \sum_{k=0}^{\infty}\frac{u^{k+1}}{k+1}$$

$$e^{-[-\log(1-u)]^{p}} = \sum_{h=0}^{\infty}(-1)^{h}\frac{[-\log(1-u)]^{ph}}{h!},$$

$$\frac{1}{1-u} = \sum_{i=0}^{\infty}u^{i},$$

$$[-\log(1-u)]^{p\alpha-1} = u^{p\alpha-1}\left[\sum_{j=0}^{\infty}\binom{p\alpha-1}{j}u^j\left(\sum_{s=0}^{\infty}\frac{u^s}{s+2}\right)^j\right],$$

and applying the result on the power series raised to a positive integer, with $c_s = (s+2)^{-1}$, that is

$$\left(\sum_{s=0}^{\infty}c_s u^s\right)^j = \sum_{s=0}^{\infty}a_{s,j}u^s,$$

where $a_{s,j} = (sc_0)^{-1}\sum_{l=1}^{s}[j(l+1)-s]c_l a_{s-l,j}$, and $a_{0,j} = c_0^j$, (Gradshteyn and Ryzhik, 2000), the GGBIII pdf can be written as

$$\begin{aligned}
g_{GGBIII}(x) &= \frac{p\lambda^\delta\beta\delta x^{-\delta-1}\left[1+\left(\frac{x}{\lambda}\right)^{-\delta}\right]^{-\beta-1}}{\Gamma(\alpha)}u^{p\alpha-1}\\
&\times \sum_{j=0}^{\infty}\sum_{s=0}^{\infty}\binom{p\alpha-1}{j}a_{s,j}u^{j+s}\sum_{h=0}^{\infty}(-1)^h\frac{[-\log(1-u)]^{ph}}{h!}\sum_{i=0}^{\infty}u^i
\end{aligned} \quad (21)$$

$$\begin{aligned}
&= \frac{p\lambda^\delta\beta\delta x^{-\delta-1}\left[1+\left(\frac{x}{\lambda}\right)^{-\delta}\right]^{-\beta-1}}{\Gamma(\alpha)}\\
&\times \sum_{j=0}^{\infty}\sum_{s,=0}^{\infty}\sum_{h=0}^{\infty}\sum_{i=0}^{\infty}\binom{p\alpha-1}{j}(-1)^h\frac{[-\log(1-u)]^{ph}}{h!}a_{s,j}u^{p\alpha+s+j+i-1}
\end{aligned} \quad (22)$$

$$\begin{aligned}
&= \sum_{j=0}^{\infty}\sum_{s=0}^{\infty}\sum_{h=0}^{\infty}\sum_{i=0}^{\infty}(-1)^h\binom{p\alpha-1}{j}\frac{p\lambda^\delta\beta\delta x^{-\delta-1}\left[1+\left(\frac{x}{\lambda}\right)^{-\delta}\right]^{-\beta-1}}{\Gamma(\alpha)}\\
&\times \frac{[-\log(1-u)]^{ph}}{h!}a_{s,j}u^{p\alpha+s+j+i-1}
\end{aligned} \quad (23)$$

$$\begin{aligned}
&= \sum_{j=0}^{\infty}\sum_{s=0}^{\infty}\sum_{h=0}^{\infty}\sum_{i=0}^{\infty}(-1)^h\binom{p\alpha-1}{j}\frac{p\lambda^\delta\beta\delta x^{-\delta-1}\left[1+\left(\frac{x}{\lambda}\right)^{-\delta}\right]^{-\beta-1}}{\Gamma(\alpha)}\\
&\times \frac{[-\log(1-u)]^{ph}}{h!}a_{s,j}\left[1+\left(\frac{x}{\lambda}\right)^{-\delta}\right]^{-\beta p\alpha+\beta s+\beta j+\beta i+\beta}
\end{aligned} \quad (24)$$

$$\begin{aligned}
&= \sum_{j=0}^{\infty}\sum_{s=0}^{\infty}\sum_{h=0}^{\infty}\sum_{i=0}^{\infty}(-1)^h\binom{p\alpha-1}{j}\frac{p\lambda^\delta\beta\delta x^{-\delta-1}}{\Gamma(\alpha)}\\
&\times a_{s,j}\left[1+\left(\frac{x}{\lambda}\right)^{-\delta}\right]^{-\beta(p\alpha+s+j+i)-1}\frac{[-\log(1-u)]^{ph}}{h!}
\end{aligned} \quad (25)$$

$$= \sum_{j=0}^{\infty}\sum_{s=0}^{\infty}\sum_{h=0}^{\infty}\sum_{i=0}^{\infty} (-1)^h \binom{p\alpha-1}{j}\frac{p\lambda^\delta \beta(p\alpha+s+j+i)\delta x^{-\delta-1}}{\Gamma(\alpha)(p\alpha+s+j+i)} a_{s,j}$$
$$\times \left[1+\left(\frac{x}{\lambda}\right)^{-\delta}\right]^{-\beta(p\alpha+s+j+i)-1} \frac{[-\log(1-u)]^{ph}}{h!} \tag{26}$$

$$= \sum_{j=0}^{\infty}\sum_{s=0}^{\infty}\sum_{h=0}^{\infty}\sum_{i=0}^{\infty} \frac{(-1)^h}{h!}\binom{p\alpha-1}{j}\frac{p\lambda^\delta \beta(p\alpha+s+j+i)\delta x^{-\delta-1}}{\Gamma(\alpha)(p\alpha+s+j+i)} a_{s,j}$$
$$\times \left[1+\left(\frac{x}{\lambda}\right)^{-\delta}\right]^{-\beta(p\alpha+s+j+i)-1} \left[-\log\left(1-\left(1+\left(\frac{x}{\lambda}\right)^{-\delta}\right)^{-\beta}\right)\right]^{ph} \tag{27}$$

where $f(x; \lambda, \beta(p\alpha+j+s+i), \delta)$ is the Burr III pdf with parameters $\lambda, \beta(p\alpha+j+s+i)$ and $\delta$. Let $C = \{(j,h,s,i) \in \mathbf{Z}_+^3\}$

$$\varphi_v = \frac{(-1)^h}{h!}\binom{p\alpha-1}{j}\frac{a_{s,j}}{\Gamma(\alpha)(p\alpha+s+j+i)}, \tag{28}$$

and

$$g_{GGBIII}(x) = \sum_{v \in C} \varphi_v f(x; \lambda, \beta(p\alpha+j+s+i), \delta). \tag{29}$$

### 2.3 Hazard and Reverse Hazard functions

Let $X$ be a continuous random variable with distribution function $F$, and probability density function (pdf) $f$, then the hazard function, reverse hazard function are given by $h_F(x) = f(x)/(1-F(x))$. The hazard and reverse hazard function of the GGBIII distribution are

$$h(x) = \frac{\dfrac{p\beta\lambda^\delta \delta x^{-\delta-1}\left(1+\left(\frac{x}{\lambda}\right)^{-\delta}\right)^{-\beta-1}}{\left(1-\left(1+\left(\frac{x}{\lambda}\right)^{-\delta}\right)^{-\beta}\right)} \times \left[-\log\left(1-\left(1+\left(\frac{x}{\lambda}\right)^{-\delta}\right)^{-\beta}\right)\right]^{p\alpha-1} \times e^{-\left[-\log\left(1-\left(1+\left(\frac{x}{\lambda}\right)^{-\delta}\right)^{-\beta}\right)\right]^p}}{\Gamma(\alpha) - \gamma\left(\left[-\log\left[1-\left(1+\left(\frac{x}{\lambda}\right)^{-\delta}\right)^{-\beta}\right]\right]^p, \alpha\right)}$$

and

$$\tau(x) = \frac{\dfrac{p\beta\lambda^\delta \delta x^{-\delta-1}\left(1+\left(\frac{x}{\lambda}\right)^{-\delta}\right)^{-\beta-1}}{\left(1-\left(1+\left(\frac{x}{\lambda}\right)^{-\delta}\right)^{-\beta}\right)} \times \left[-\log\left(1-\left(1+\left(\frac{x}{\lambda}\right)^{-\delta}\right)^{-\beta}\right)\right]^{p\alpha-1} \times e^{-\left[-\log\left(1-\left(1+\left(\frac{x}{\lambda}\right)^{-\delta}\right)^{-\beta}\right)\right]^p}}{\gamma\left(\left[-\log\left[1-\left(1+\left(\frac{x}{\lambda}\right)^{-\delta}\right)^{-\beta}\right]\right]^p, \alpha\right)} \tag{30}$$

respectively, for $x \geq 0, \beta \geq 0, \delta \geq 0, \alpha \geq 0, p \geq 0,$ and $\lambda \geq 0$.

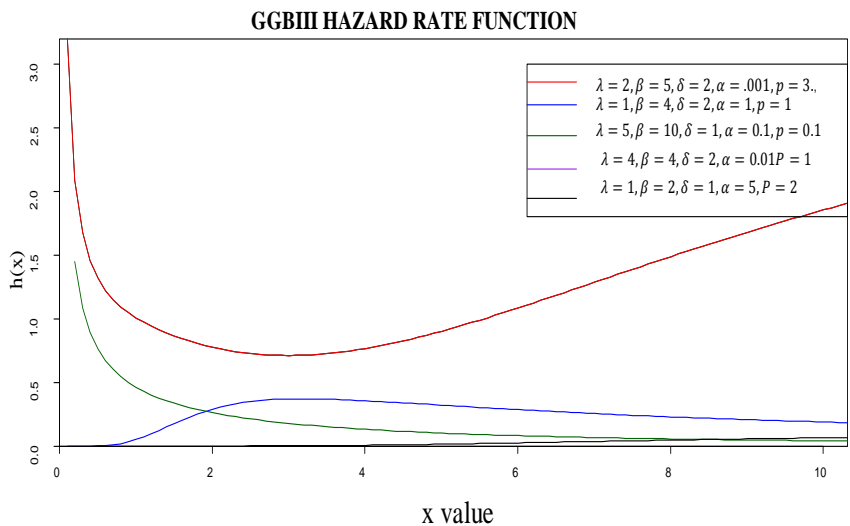

FIGURE 2: THE Plot of Hazard Rate Function of GGBIII

The plots show various shapes including monotonically decreasing, monotonically increasing, and bathtub followed by upside down bathtub shapes for five combinations of values of the parameters. This very attractive flexibility makes the GGBIII hazard rate function useful and suitable for monotonic and non-monotone empirical hazard behaviors which are more likely to be encountered or observed in real life situations.

### 2.4 GGBIII Quantile Function

Let $X$ be a random variable with distribution function $F$, and let $q \in (0,1)$. A value of $x$ such that $F(x^-) = P(x < x) \leq q$ and $F(x) = P(X \leq x) \geq q$ is called a quantile of order $q$ for the distribution. Roughly speaking, a quantile of order $q$ is a value where the graph of the cumulative distribution function crosses (or jumps over) $q$.
The quantile function of GGBIII distribution is obtained by solving the equation

Thus, the quantile function is given by

$$F^{-1}(q) = min\{x \in \mathbf{R} : F(x) \geq q\}, \quad q \in (0,1)$$

$$Q_{GGBIII}(x_p) = q, \quad 0 < q < 1.$$

$$Q_{GGBIII}(q) = \lambda \left[\left(1 - e^{-u^{1/p}}\right)^{-1/\beta} - 1\right]^{-1/\delta} \tag{31}$$

then $\gamma(u, \alpha) = q\Gamma(\alpha)$, where $u = \gamma^{-1}(q\Gamma(\alpha), \alpha)$.

## 3     MOMENTS, MOMENT GENERATING FUNCTION, MEAN AND MEDIAN DEVIATIONS

In this section, we present the moments, moment generating function, mean and median deviations for the GGBIII distribution.

### 3.1     Moments and Moment Generating Function

As with any other distribution, many of the interesting characteristics and features of the GGBIII distribution can be studied through the moments. Let $\beta^* = \beta(p\alpha + j + s + i)$, and $Y \sim BIII(\delta, \lambda, \beta^*)$. Then the $r^{th}$ moment of the random variable $Y$ is

$$E(Y^r) = \beta^* \lambda^r B\left(\beta^* + \frac{r}{\delta}, 1 - \frac{r}{\delta}\right), \quad r < \delta \tag{32}$$

So that the $r^{th}$ raw moment is thus given by the following

$$E(X^r) = \sum_{v \in C} \varphi_v \beta^* \lambda^r B\left(\beta^* + \frac{r}{\delta}, 1 - \frac{r}{\delta}\right) \quad r < \delta, \tag{33}$$

The moment generating function of the BGBIII distribution is given by

$$M_X(t) = E(e^{tX}) = E\left[1 + tX + \frac{(tX)^2}{2!} + \frac{(tX)^3}{3!} + \cdots\right], \tag{34}$$

$$= \sum_{r=1}^{\infty} \frac{t^r}{r!} E(X^r),$$

then we have,

$$\sum_{r=0}^{\infty} \sum_{v \in C} \frac{t^r}{r!} \varphi_v \beta^* \lambda^r B(\beta^* + r/\delta, 1 - r/\delta) \quad r < \delta, \tag{35}$$

### 3.2     Mean and Median Deviations

#### 3.2.1     Mean Deviation

If X has the GGBIII distribution, we derive the mean deviation about the mean $\mu$ by

$$\delta_1 = \int_0^\infty |x - \mu| g_{GD}(x) dx = 2\mu G_{GGBIII}(\mu) - 2\mu + 2T(\mu), \tag{36}$$

Where $\mu = E(X)$ and $T(\mu) = \int_\mu^\infty x \cdot g_{GGBIII}(x) dx$. Let $\beta^* = \beta(p\alpha + j + s + i)$, then

$$T(\mu) = \sum_{v \in C} \varphi_v T_{GGBIII(\beta^*, \delta, \lambda)}(\mu) \tag{37}$$

$$= \sum_{v \epsilon C} \varphi_v \beta^* \lambda [B(\beta^* + 1/\delta, 1 - 1/\delta) - B(t(\mu); \beta^* + 1/\delta, 1 - 1/\delta)] \qquad (38)$$

$$\delta_1 = 2 \sum_{v \epsilon C} \varphi_v \beta^* \lambda [B(\beta^* + 1/\delta, 1 - 1/\delta)] \left[ \frac{\gamma\left(\left[-\log\left[1-\left(1+\left(\frac{\mu}{\lambda}\right)^{-\delta}\right)^{-\beta}\right]\right]^p, \alpha\right)}{\Gamma(\alpha)} - 1 \right]$$

$$+ 2 \sum_{v \epsilon C} \varphi_v \beta^* \lambda [B(\beta^* + 1/\delta, 1 - 1/\delta) - B(t(\mu); \beta^* + 1/\delta, 1 - 1/\delta)] \qquad (39)$$

Where $t(\mu) = \left(1 + \left(\frac{\mu}{\lambda}\right)^{-\delta}\right)^{-1}$ and $B(x; a, b) = \int_0^x t^{a-1}(1-t)^{b-1} dt$.

### 3.2.2 Median Deviation

If $X$ has the GGBIII distribution, we derive the median deviation about the median $M$ by

$$\delta_2 = \int_0^\infty |x - M| g_{GD}(x) dx = 2T(M) - \mu, \qquad (40)$$

where $T(M) = \int_M^\infty x \cdot g_{GGBIII}(x) dx$

$$M = Q_{GGBIII}(0.5) = \lambda \left[ \left(1 - e^{-[\gamma^{-1}(0.5\Gamma(\alpha),\alpha)]^{1/p}}\right)^{-1/\beta} - 1 \right]^{-1/\delta} \qquad (41)$$

$$T(M) = \sum_{v \epsilon C} \varphi_v T_{BIII(\beta^*, \delta, \lambda)}(M)$$

$$= \sum_{v \epsilon C} \varphi_v \beta^* \lambda [B(\beta^* + 1/\delta, 1 - 1/\delta) - B(t(M); \beta^* + 1/\delta, 1 - 1/\delta)] \qquad (42)$$

$$2 \sum_{v \epsilon C} \varphi_v \beta^* \lambda [B(\beta^* + 1/\delta, 1 - 1/\delta) - B(t(M); \beta^* + 1/\delta, 1 - 1/\delta)]$$

$$- \sum_{v \epsilon C} \varphi_v \beta^* \lambda [B(\beta^* + 1/\delta, 1 - 1/\delta)] \qquad (43)$$

### 4 ORDER STATISTICS OF GGBIII DISTRIBUTION

Definition: Let $(y_1, ..., y_n)$ be a random sample from the GGBIII distribution with pdf f(y) defined over the interval $-\infty$ to $\infty$. A rearrangement of the random sample into $(x_1, ..., x_n)$ i.e. $-\infty < x_1, x_2, ... x_n < \infty$ is known as the ordered transformation of the random sample and $x_1, x_2, ..., x_n$ are called ordered statistics.

Theorem: Let $x_1, x_2, ..., x_n$ be ordered statistics of a random sample $(y_1, ..., y_n)$ from a GGBIII distribution with pdf $g(x)$, the joint pdf is given as

$$f(x_1, x_2, ..., x_n) = n! \, g(x_1) g(x_2) ... g(x_n) \, for -\infty < x_1 < x_2 < \cdots < x_n < \infty.$$

The general formula for order statistics is given by

$$f_{i;n}(x) = \frac{n!\, g(x)}{(n-i)!\,(i-1)!} [G(x)]^{i-1}[1-G(x)]^{n-i}, \tag{44}$$

Again, using the binomial expansion to the second factor we obtain

$$f_{i;n}(x) = \frac{n!\, g(x)}{(n-i)!\,(i-1)!} \sum_{j=0}^{n-i} (-1)^j \binom{n-i}{j} [G(x)]^{i+j-1}, \tag{45}$$

$$f_{i;n}(x) = \frac{n!\, g(x)}{(n-i)!\,(i-1)!} \sum_{j=0}^{n-i} (-1)^j \binom{n-i}{j} \left[\frac{\gamma([-\log(1-F(x))]^p, \alpha)}{\Gamma(\alpha)}\right]^{i+j-1} \tag{46}$$

From Gratyshetny and Ryhzik, 2000

$$\gamma(x, \alpha) = \sum_{k=0}^{\infty} \frac{(-1)^k x^{k+\alpha}}{(k+\alpha)k!},$$

and if $d_k = (-1)^k/((k+\alpha)k!)$, then

$$\begin{aligned}f_{i;n}(x) &= \frac{n!\, g(x)}{(n-i)!\,(i-1)!} \sum_{j=0}^{n-i} \binom{n-i}{j} \frac{(-1)^j}{[\Gamma(\alpha)]^{i+j-1}} \left([-\log(1-F(x))]^p\right)^{\alpha(i+j-1)} \\ &\quad \times \left[\sum_{k=0}^{\infty} \frac{(-1)^k ([[-\log(1-F(x))]^p])^k}{(k+\alpha)k!}\right]^{i+j-1}\end{aligned} \tag{47}$$

$$\begin{aligned}f_{i;n}(x) &= \frac{n!\, g(x)}{(n-i)!\,(i-1)!} \sum_{j=0}^{n-i} \binom{n-i}{j} \frac{(-1)^j}{[\Gamma(\alpha)]^{i+j-1}} \left([-\log(1-F(x))]^p\right)^{\alpha(i+j-1)} \\ &\quad \times \sum_{k=0}^{\infty} \left([-\log(1-F(x))]^p\right)^k,\end{aligned} \tag{49}$$

$$\begin{aligned}f_{i;n}(x) &= \frac{n!\, g(x)}{(n-i)!\,(i-1)!} \sum_{j=0}^{n-i}\sum_{k=0}^{\infty} \binom{n-i}{j} \frac{(-1)^j d_{k,n-i+j}}{[\Gamma(\alpha)]^{i+j-1}} \left([-\log(1-F(x))]^p\right)^{\alpha(i+j-1)+k}\end{aligned} \tag{50}$$

$$\begin{aligned}&= \frac{n!\, p([-\log(1-F(x))])^{p\alpha-1} e^{-[-\log(1-F(x))]^p} [1-F(x)]^{-1} f(x)}{(n-i)!\,(i-1)!\,\Gamma(\alpha)} \sum_{j=0}^{n-i}\sum_{k=0}^{\infty} \binom{n-i}{j} \frac{(-1)^j d_{m,i+j-1}}{[\Gamma(\alpha)]^{i+j-1}} \\ &\quad \times \left([-\log(1-F(x))]^p\right)^{\alpha(i+j-1)+k}\end{aligned} \tag{51}$$

$$\begin{aligned}&= \frac{n!}{(n-i)!\,(i-1)!} \sum_{j=0}^{n-i}\sum_{k=0}^{\infty} \binom{n-i}{j} \frac{(-1)^j d_{m,i+j-1}}{[\Gamma(\alpha)]^{i+j}} ([-\log(1-F(x))]^p)^{p\alpha(i+j)+k-1} e^{-[-\log(1-F(x))]^p} [1-F(x)]^{-1} f(x)\end{aligned} \tag{52}$$

$$\begin{aligned}&= \frac{n!}{(n-i)!\,(i-1)!} \sum_{j=0}^{n-i}\sum_{k=0}^{\infty} \binom{n-i}{j} \frac{(-1)^j d_{m,i+j-1}}{[\Gamma(\alpha)]^{i+j}} \times \frac{\Gamma((\alpha(i+j)+k))}{\Gamma((\alpha(i+j)+k))} [-\log(1-F(x))]^{p\alpha(i+j-1)+k-1} \\ &\quad \times e^{-[-\log(1-F(x))]^p} \times [1-F(x)]^{-1} f(x)\end{aligned} \tag{53}$$

That is,

$$f_{i,n}(x) = \frac{n!}{(n-i)!(i-1)!} \sum_{j=0}^{n-i} \sum_{k=0}^{\infty} \binom{n-i}{j} \frac{(-1)^j d_{m,i+j-1} \Gamma(\alpha(i+j)+k)}{[\Gamma(\alpha)]^{i+j}}$$
$$\times f(x; \alpha(i+j)+k, p, \beta, \lambda, \delta), \tag{54}$$

where $f(x; \alpha(i+j)+k, p, \beta, \lambda, \delta)$ is the GGBIII pdf with parameters $\beta, \delta, \lambda, p$ and shape parameter $\alpha^* = \alpha(i+j)+k$.

$$= \frac{n!}{(n-i)!(i-1)!} \sum_{j=0}^{n-i} \sum_{k=0}^{\infty} \binom{n-i}{j} \frac{(-1)^j d_{m,i+j-1} \Gamma((\alpha(i+j)+k))}{[\Gamma(\alpha)]^{i+j}} \frac{p\beta\delta\lambda^\delta x^{-\delta-1}\left(1+\left(\frac{x}{\lambda}\right)^{-\delta}\right)^{-\beta-1}}{\Gamma((\alpha(i+j)+k))} \left[-\log\left(1\right.\right.$$
$$\left.-\left(1+\left(\frac{x}{\lambda}\right)^{-\delta}\right)^{-\beta}\right)\right]^{p\alpha(i+j-1)+k-1} e^{-\left[-\log\left(1-\left(1+\left(\frac{x}{\lambda}\right)^{-\delta}\right)^{-\beta}\right)\right]^p} \left[1-\left(1+\left(\frac{x}{\lambda}\right)^{-\delta}\right)^{-\beta}\right]^{-1} \tag{55}$$

## 5    MAXIMUM LIKELIHOOD ESTIMATION

The maximum likelihood estimation (MLE) is one of the most widely used estimation method for finding the unknown parameters. Consider a random sample $x_1, x_2, \ldots, x_n$ from the generalized gamma-Dagum distribution.
The likelihood function is given by

$$L(\beta, \alpha, \delta, \lambda, p) = \frac{(p\beta\lambda^\delta\delta)^n}{[\Gamma(\alpha)]^n} \prod_{i=1}^{n} \left\{ x_i^{-\delta-1}\left[1+\left(\frac{x_i}{\lambda}\right)^{-\delta}\right]^{-\beta-1} \left[-\log\left(1-\left(1+\left(\frac{x_i}{\lambda}\right)^{-\delta}\right)^{-\beta}\right)\right]^{p\alpha-1} \left[1 - (1+\lambda x_i^{-\delta})^{-\beta}\right]^{-1} \times e^{-\left[-\log\left(1-\left(1+\left(\frac{x_i}{\lambda}\right)^{-\delta}\right)^{-\beta}\right)\right]^p} \right\} \tag{56}$$

where $\Theta = (p, \alpha, \beta, \delta, \lambda)^T$

Now, the log-likelihood function denoted by $\ell$

$\ell = \log[L(\Theta)]$

$L(\theta) = n\log(\lambda^\delta) + n\log(p) + n\log(\beta) + n\log(\delta) - n\log\Gamma(\alpha) - (\delta+1)\sum_{i=1}^{n}\log(x_i) - (\beta+1)\sum_{i=1}^{n}\log\left(1+\left(\frac{x_i}{\lambda}\right)^{-\delta}\right) + (p\alpha-1)\sum_{i=1}^{n}\log\left[-\log\left(1-(1+\lambda x_i^{-\delta})^{-\beta}\right)\right] - \sum_{i=0}^{n}\log\left[1-(1+\lambda x_i^{-\delta})^{-\beta}\right] - \sum_{i=1}^{n}\left[-\log\left(1-(1+\lambda x_i^{-\delta})^{-\beta}\right)\right]^p \tag{57}$

The partial derivatives of $\ell$ with respect to the parameters are

$$\frac{\partial l}{\partial \alpha} = -n\frac{\Gamma'(\alpha)}{\Gamma(\alpha)} + p\sum_{i=0}^{n}\log\left[-\log\left(1-\left(1+\left(\frac{x_i}{\lambda}\right)^{-\delta}\right)^{-\beta}\right)\right] \tag{58}$$

$$\frac{\partial l}{\partial \beta} = \frac{n}{\beta} - \sum_{i=1}^{n} \log\left(1+\left(\frac{x_i}{\lambda}\right)^{-\delta}\right) - (p\alpha-1)\sum_{i=0}^{n} \frac{\left(1+\left(\frac{x_i}{\lambda}\right)^{-\delta}\right)^{-\beta} \log\left(1+\left(\frac{x_i}{\lambda}\right)^{-\delta}\right)}{\left(1-\left(1+\left(\frac{x_i}{\lambda}\right)^{-\delta}\right)^{-\beta}\right)\log\left(1-\left(1+\left(\frac{x_i}{\lambda}\right)^{-\delta}\right)^{-\beta}\right)}$$

$$-\sum_{i=0}^{n} \frac{\left(1+\left(\frac{x_i}{\lambda}\right)^{-\delta}\right)^{-\beta} \log\left(1+\left(\frac{x_i}{\lambda}\right)^{-\delta}\right)}{\left(1-\left(1+\left(\frac{x_i}{\lambda}\right)^{-\delta}\right)^{-\beta}\right)}$$

$$-p\sum_{i=1}^{n} \frac{\left[-\log\left(1-\left(1+\left(\frac{x_i}{\lambda}\right)^{-\delta}\right)^{-\beta}\right)\right]^{p-1} \left(1+\left(\frac{x_i}{\lambda}\right)^{-\delta}\right)^{-\beta} \log\left(1+\left(\frac{x_i}{\lambda}\right)^{-\delta}\right)}{\left(1-\left(1+\left(\frac{x_i}{\lambda}\right)^{-\delta}\right)^{-\beta}\right)} \qquad (59)$$

$$\frac{\partial l}{\partial \delta} = \frac{n}{\delta} - n\log\lambda - \sum_{i=1}^{n} \log x_i + (\beta+1)\sum_{i=1}^{n} \frac{\lambda x_i^{-\delta} \log x_i}{1+\left(\frac{x^i}{\lambda}\right)^{-\delta}}$$

$$+ (p\alpha-1)\sum_{i=1}^{n} \frac{\left(1+\left(\frac{x_i}{\lambda}\right)^{-\delta}\right)^{-\beta} \left(\frac{x_i}{\lambda}\right)^{-\delta} \log\left(\frac{x_i}{\lambda}\right)}{\left[1-\left(1+\left(\frac{x_i}{\lambda}\right)^{-\delta}\right)^{-\beta}\right]\left(\log\left(1-\left(1+\left(\frac{x_i}{\lambda}\right)^{-\delta}\right)^{-\beta}\right)\right)}$$

$$-p\sum_{i=1}^{n} \frac{\left[-\log\left(1-\left(1+\left(\frac{x_i}{\lambda}\right)^{-\delta}\right)^{-\beta}\right)\right]^{p-1} \left(\frac{x_i}{\lambda}\right)^{-\delta} \left(1+\left(\frac{x_i}{\lambda}\right)^{-\delta}\right)^{-\beta} \log\left(\frac{x_i}{\lambda}\right)}{\left[1-\left(1+\left(\frac{x_i}{\lambda}\right)^{-\delta}\right)^{-\beta}\right]}$$

$$+\sum_{i=1}^{n} \frac{\left(1+\left(\frac{x_i}{\lambda}\right)^{-\delta}\right)^{-\beta} \left(\frac{x_i}{\lambda}\right)^{-\delta} \log\left(\frac{x_i}{\lambda}\right)}{\left(1-\left(1+\left(\frac{x_i}{\lambda}\right)^{-\delta}\right)^{-\beta}\right)} \qquad (60)$$

$$\frac{\partial l}{\partial \lambda} = \frac{n\delta}{\lambda} - (\beta+1)\sum_{i=1}^{n} \frac{\left(\frac{x_i}{\lambda}\right)^{-\delta}\frac{\delta}{\lambda}}{1+\left(\frac{x_i}{\lambda}\right)^{-\delta}} + (p\alpha-1)\sum_{i=1}^{n} \frac{\beta\frac{\delta}{\lambda}\left(\frac{x_i}{\lambda}\right)^{-\delta}(1+\lambda x_i^{-\delta})^{-\beta}}{\left(1-\left(1+\left(\frac{x_i}{\lambda}\right)^{-\delta}\right)^{-\beta}\right)\left[-\log\left(1-\left(1+\left(\frac{x_i}{\lambda}\right)^{-\delta}\right)^{-\beta}\right)\right]}$$

$$+p\sum_{i=1}^{n} \frac{\left[-\log\left(1-\left(1+\left(\frac{x_i}{\lambda}\right)^{-\delta}\right)^{-\beta}\right)\right]^{p-1} \left(1+\left(\frac{x_i}{\lambda}\right)^{-\delta}\right)^{-\beta} \left(\frac{x_i}{\lambda}\right)^{-\delta}\frac{\delta}{\lambda}}{\left[1-\left(1+\left(\frac{x_i}{\lambda}\right)^{-\delta}\right)^{-\beta}\right]}$$

$$-\sum_{i=1}^{n} \frac{(1+\lambda x_i^{-\delta})^{-\beta} \left(\frac{x_i}{\lambda}\right)^{-\delta}\frac{\delta}{\lambda}}{\left[1-\left(1+\left(\frac{x_i}{\lambda}\right)^{-\delta}\right)^{-\beta}\right]} \qquad (61)$$

$$\frac{\partial l}{\partial p} = \frac{n}{p} + \alpha \sum_{i=0}^{n} \log\left[-\log\left[1 - \left(1 + \left(\frac{x_i}{\lambda}\right)^{-\delta}\right)^{-\beta}\right]\right]$$
$$+ \sum_{i=1}^{n} \left(\log\left[1 - \left(1 + \left(\frac{x_i}{\lambda}\right)^{-\delta}\right)^{-\beta}\right]\right)^p \log\left(\log\left[1 - \left(1 + \left(\frac{x_i}{\lambda}\right)^{-\delta}\right)^{-\beta}\right]\right) \quad (62)$$

respectively. The MLE of the parameters $\beta, \delta, \lambda, p, \alpha$ will be obtained and they are denoted by $\hat{\beta}, \hat{\lambda}, \hat{p}, \hat{\delta}, \hat{\alpha}$.

### 5.1 Asymptotic confidence intervals

The asymptotic confidence intervals for the parameters of the GGBIII distribution are presented. The expectations in the Fisher Information Matrix (FIM) will now be obtained numerically. The approximate $100(1 - \eta)\%$ two-sided confidence intervals for $\beta, \delta, \lambda, \alpha, p$. are given by:

$$\hat{\beta} \pm Z_{\frac{n}{2}}\sqrt{I_{\beta\beta}^{-1}(\widehat{\Theta})}, \quad \hat{\lambda} \pm Z_{\frac{n}{2}}\sqrt{I_{\lambda\lambda}^{-1}(\widehat{\Theta})}, \quad \hat{\delta} \pm Z_{\frac{n}{2}}\sqrt{I_{\delta\delta}^{-1}(\widehat{\Theta})}, \quad \hat{\alpha} \pm Z_{\frac{n}{2}}\sqrt{I_{\alpha\alpha}^{-1}(\widehat{\Theta})}, \quad \hat{p} \pm Z_{\frac{n}{2}}\sqrt{I_{pp}^{-1}(\widehat{\Theta})},$$

Under the usual regularity conditions, the well-known asymptotic properties of the maximum likelihood method ensure that $\sqrt{n}\left(\widehat{\Theta_n} - \Theta\right) \xrightarrow{d} N(0, \Sigma_\Theta)$, where the $\Sigma_\Theta = [I(\Theta)]^{-1}$ is the asymptotic variance-covariance matrix and

$$I(\Theta) = \begin{pmatrix} J_{\alpha\alpha} & J_{\beta\alpha} & J_{\delta\alpha} & J_{\lambda\alpha} & J_{p\alpha} \\ J_{\alpha\beta} & J_{\beta\beta} & J_{\delta\beta} & J_{\lambda\beta} & J_{p\beta} \\ J_{\alpha\delta} & J_{\beta\delta} & J_{\delta\delta} & J_{\lambda\delta} & J_{p\delta} \\ J_{\alpha\lambda} & J_{\beta\lambda} & J_{\delta\lambda} & J_{\lambda\lambda} & J_{p\lambda} \\ J_{\alpha p} & J_{\beta p} & J_{\delta p} & J_{\lambda p} & J_{pp} \end{pmatrix}$$

We can use the likelihood ratio (LR) test to compare the fit of the GD distribution with its sub-models for a given data set. The LR test rejects the null hypothesis if $\omega > \chi_c^2$ where $\chi_c^2$ denote the upper 100% point of the $\chi^2$ distribution with 2 degrees of freedom.

### 6    APPLICATIONS

In this section, we present applications of the proposed GGBIII distribution and compare it to the existing Gamma-G families in real data sets to illustrate its potentiality and robustness. The maximum likelihood estimates (MLEs) of the GGBIII parameters $\alpha, \beta, \delta, \lambda$ and $p$ are computed by maximizing the objective function via the sub-routine NLMIXED in SAS. The estimated values of the parameters (standard error in parenthesis), -2log-likelihood statistic, Akaike Information Criterion, Bayesian Information Criterion, and Corrected Akaike Information Criterion are presented. Also, presented are values of Likelihood Ratio test, Kolmogorov-Smirnov, Cramer-Von Mises, Anderson-Darling statistic for hypothesis test which were obtained using a package fitdistrplus in R. In order to compare the models above with the proposed model, we applied formal goodness-of-fit tests to verify which distribution fits better to the real data sets. Here, we consider the Anderson-Darling $(A)$, Cramér-von Mises $(W)$ and Kolmogorov-Smirnov statistics $KS$. In general, the distribution which has the smallest values of these statistics is the better fit for the data.

Table 1: Descriptive Statistics for the Datasets Used

| Data | N | Mean | Median | Mode | SD | Var | Skewness | Kurtosis | Min | Max |
|---|---|---|---|---|---|---|---|---|---|---|
| Myelo | 33 | 40.88 | 22.00 | 4.00 | 46.70 | 2181.17 | 1.22 | 0.35 | 1 | 156 |
| Aircon | 188 | 92.07 | 54.00 | 14.00 | 107.92 | 11645.93 | 2.16 | 5.19 | 1 | 603 |
| Airc | 30 | 85.93 | 22.00 | 11.00 | 165.72 | 27463.58 | 4.06 | 18.83 | 1 | 877 |
| Comp | 50 | 3.34 | 1.41 | | 4.181 | 17.48 | 1.46 | 1.33 | 0.04 | 15.04 |
| Cancer | 1207 | 46.96 | 42.97 | 18.67 | 29.63 | 878.46 | 0.63 | - 0.24 | 2.63 | 133.80 |

### 6.1 Acute Myelogeneous Leukemia

The first real data set represents the survival times, in weeks, of 33 patients suffering from acute Myelogeneous Leukemia. These data have been analyzed by Feigl and Zelen (1965). The data are: 65, 156, 100, 134, 16, 108, 121, 4, 39, 143, 56, 26, 22, 1, 1, 5, 65, 56, 65, 17, 7, 16, 22, 3, 4, 2, 3, 8, 4, 3, 30, 4, 43. For these data, we shall compare the proposed GGBIII distribution to Gamma Dagum (GD) (Broderick *et al.*, 2014), alternative Gamma Dagum (GD) (Jailson and Ana, 2015), Zografos and Balakrishnan Dagum (ZB-D) (Zografos and Balakrishnan, 2009).

The asymptotic covariance matrix of the MLEs of the GGBIII model parameters, which is the inverse of the observed Fisher information matrix $I_n^{-1}(\widehat{\Delta})$ is given by:

$$= \begin{pmatrix} 0.0046209593 & 0.0106779546 & -0.000327988 & 2.427039e-04 & 12.26055684 \\ 0.0106779546 & 0.0246161742 & -0.000759203 & 4.654929e-04 & 25.15292231 \\ -0.00032799 & -0.000759203 & 0.0000231093 & -5.77747e-05 & -1.55199388 \\ 0.0002427039 & 0.0004654929 & -0.000057774 & 1.899781e-07 & 0.014450710 \\ 12.260556843 & 25.152922308 & -1.551993882 & 1.445071e-02 & 1166.046687 \end{pmatrix}$$

Table 1 lists the MLEs of the model parameters for GGBIII, GD, RBD, ZBD models, the corresponding standard errors (given in parentheses) and the statistics $A$, KS and $W$.

These results show that the GGBIII distribution has the lowest $A^*$, KS and $W^*$. values among all the fitted models, and so it could be chosen as the best model.

Table 2: The Maximum Likelihood Estimation of the Generalized Gamma Burr III distribution for the Acute Myelogeneous Data

| MODEL | $\alpha$ | $\beta$ | $\lambda$ | $\delta$ | $p$ | $\theta$ |
|---|---|---|---|---|---|---|
| **GGBIII** | 0.0220 (0.0211) | 12.510 (12.5082) | 0.5962 (1.455) | 0.5817 (0.15417) | 22.09 (19.3) | – |
| **GD** | 0.1663 (0.073) | 20.1665 (3.1281) | 5.1135 (1.5994) | 0.494 (0.0252) | – | 0.001 (0.0005) |
| **ZB-D** | 24.167 (0.008) | 0.00565 (0.01492) | 0.0004 (0.0013) | 3.1276 (0.1875) | – | – |
| **RBD** | 36.585 (223.70) | 17.7768 (118.129) | 14.09 (69.47) | 0.7687 (0.0615) | – | – |

Table 3: THE AIC, AICc and BIC for the Distributions

| MODEL | $-2\text{Log}-\text{Likelihood}$ | AIC | AICC | BIC |
|---|---|---|---|---|
| **GGBIII** | 299.2 | 309.2 | 311.4 | 316.7 |
| **GD** | 303.6 | 313.6 | 315.8 | 321.1 |
| **ZB-D** | 308.7 | 316.7 | 318.2 | 322.7 |
| **RBD** | 307.4 | 315.4 | 316.8 | 321.4 |

Table 4: Likelihood Ratio Test Statistic

| MODEL | *Hypothesis* | *LR Statistic* |
|---|---|---|
| **GGBIII vs GD** | $H_0: GD \ vs \ H_1: GGBIII$ | 4.4 |
| **GGBIII vs RBD** | $H_0: RBD \ vs \ H_1: GGBIII$ | 8.2 |
| **GGBIII vs ZB-D** | $H_0: ZBD \ vs \ H_1: GGBIII$ | 9.5 |

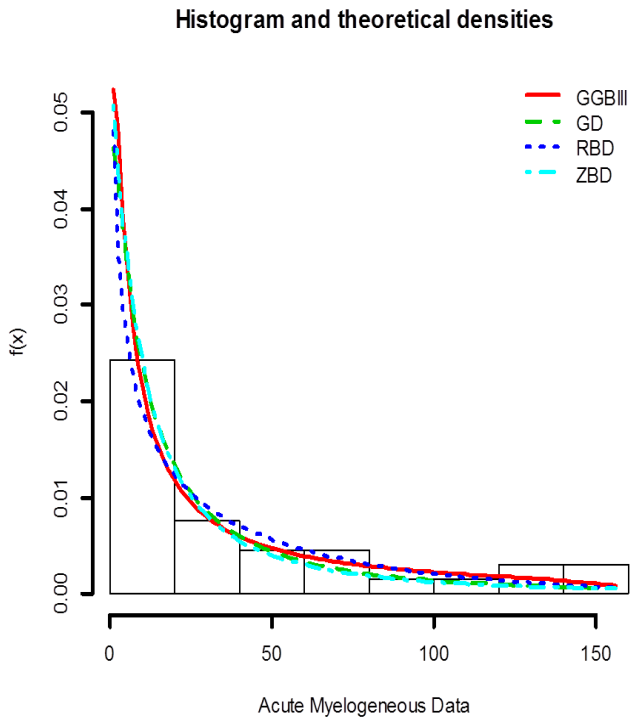
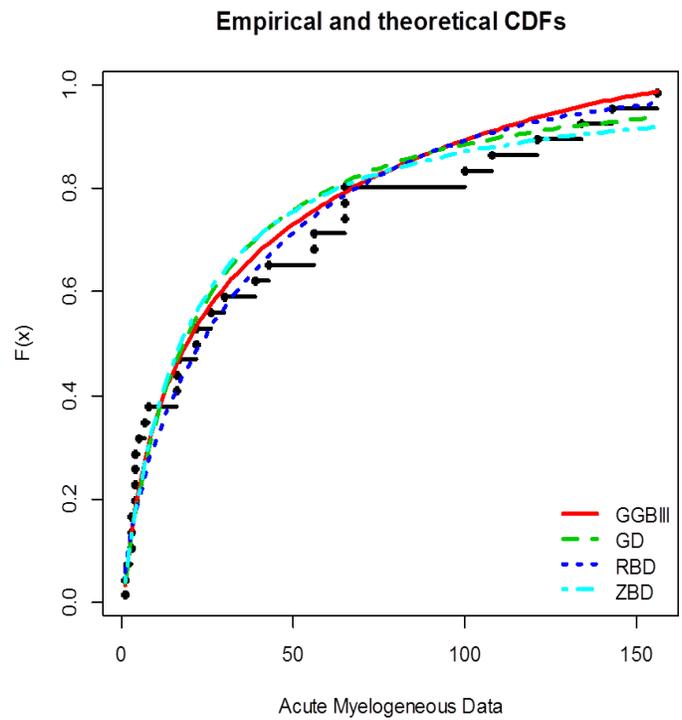
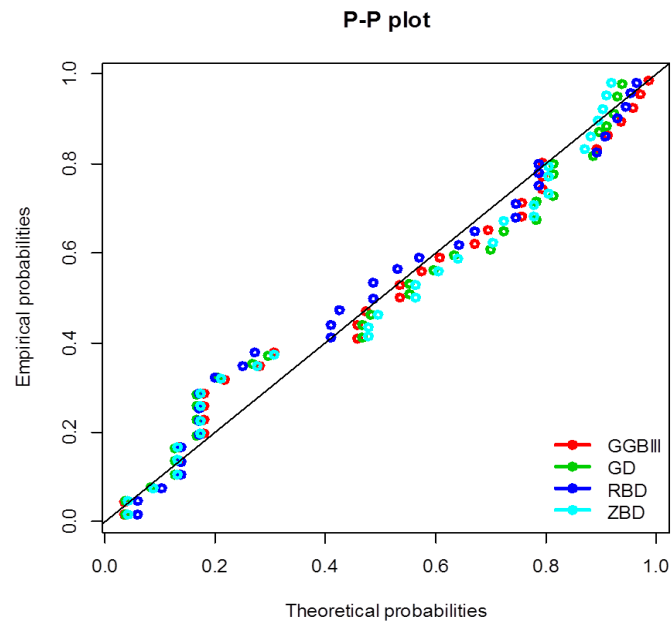

Figure 3: The histogram of the acute myelogeneous data and the estimated fitted distributions
Figure 4: The ECF of the acute myelogeneous data and the estimated fitted distributions
Figure 5: The P-P of the acute myelogeneous data and the estimated fitted distributions

## 6.2 The Air conditioning System data

The second example consists of the number of successive failures for the Air conditioning system of each member in a fleet of 13 Boeing 720 jet airplanes (Proschan, 1963).

**Table 5: Air conditioning system data**

```
194 413  90  74  55  23  97  50 359  50 130 487  57 102  15  14  10  57 320
261  51  44   9 254 493  33  18 209  41  58  60  48  56  87  11 102  12   5  14  14
 29  37 186  29 104   7   4  72 270 283   7  61 100  61 502 220 120 141  22
603  35  98  54 100  11 181  65  49  12 239  14  18  39   3  12   5  32   9 438  43
134 184  20 386 182  71  80 188 230 152   5  36  79  59  33 246   1  79   3  27
201  84  27 156  21  16  88 130  14 118  44  15  42 106  46 230  26  59 153
104  20 206   5  66  34  29  26  35   5  82  31 118 326  12  54  36  34  18  25 120
 31  22  18 216 139  67 310   3  46 210  57  76  14 111  97  62  39  30   7  44  11
 63  23  22  23  14  18  13  34  16  18 130  90 163 208   1  24  70  16 101  52
208 95  62  11 191  14  71
```

The asymptotic covariance matrix of the MLEs of the GGBIII model parameters, which is the inverse of the observed Fisher information matrix $I_n^{-1}(\hat{\Delta})$ is given by:

$$= \begin{pmatrix} 0.554348211 & 0.373214023 & -0.0199769930 & 0.0080488570 & 0.272517683 \\ 0.373214023 & 0.232368967 & -0.015293134 & 0.0046513590 & 0.151946115 \\ -0.01997699 & -0.01529313 & 0.00045638670 & -0.000727357 & -0.08481367 \\ 0.008048857 & 0.004651359 & -0.0007273570 & 6.844605e-05 & 0.001974827 \\ 0.272517683 & 0.151946115 & -0.0848136726 & 1.974827e-03 & 0.051271694 \end{pmatrix}$$

**Table 6: The Maximum Likelihood Estimation of the Generalized Gamma Burr III Distribution for the Air Conditioning System Data**

| MODEL | $\alpha$ | $\beta$ | $\lambda$ | $\delta$ | $p$ | $\theta$ |
|---|---|---|---|---|---|---|
| **GGBIII** | 0.0440 (0.0223) | 9.1609 (1.90343) | 30.8940 (13.0829) | 1.0662 (0.0993) | 3.8490 (1.8963) | − |
| **GD** | 0.1856 (0.0188) | 31.0783 (7.1966) | 2.1816 (0.8567) | 0.538 (0.05068) | − | 0.5384 (0.034) |
| **ZB-D** | 10.6110 (1.9869) | 14.8939 (1.0488) | 0.9507 (0.0375) | 0.1885 (0.01895) | − | − |
| **RBD** | 23.5930 (63.608) | 9.0448 (20.2853) | 77.9930 (124.22) | 0.4739 (0.4477) | − | − |

**Table 7: The $-2L$, AIC, AICc and BIC of the Distributions**

| MODEL | $-2$LL | AIC | AICC | BIC |
|---|---|---|---|---|
| **GGBIII** | 2062.9 | 2072.9 | 2073.3 | 2089.1 |
| **GD** | 2065.1 | 2075.1 | 2075.4 | 2091.2 |
| **ZB-D** | 2084.7 | 2092.7 | 2092.9 | 2105.7 |
| **RBD** | 2066.9 | 2074.9 | 2075.1 | 2087.8 |

### Table 8: The Likelihood Ratio Test Statistic

| MODEL | Hypothesis | LR Statistic |
|---|---|---|
| **GGBIII vs GD** | $H_0: GD \ vs \ H_1: GGBIII$ | 2.2 |
| **GGBIII vs RBD** | $H_0: GD \ vs \ H_1: GGBIII$ | 4.0 |
| **GGBIII vs ZB-D** | $H_0: GD \ vs \ H_1: GGBIII$ | 21.8 |

### Table 9: Goodness of Fit Statistic for Air Conditioning System Data

| MODEL | *Cramer−Von Mises Statistic* | *Anderson−darling Statistic* | *Kolmogorov Smirnov Statistic* |
|---|---|---|---|
| **GGBIII** | 0.03536 | 0.26363 | 0.03815 |
| **GD** | 0.04063 | 0.28902 | 0.04296 |
| **RB-D** | 0.03807 | 0.30803 | 0.04151 |
| **ZBD** | 0.13285 | 1.08089 | 0.05574 |

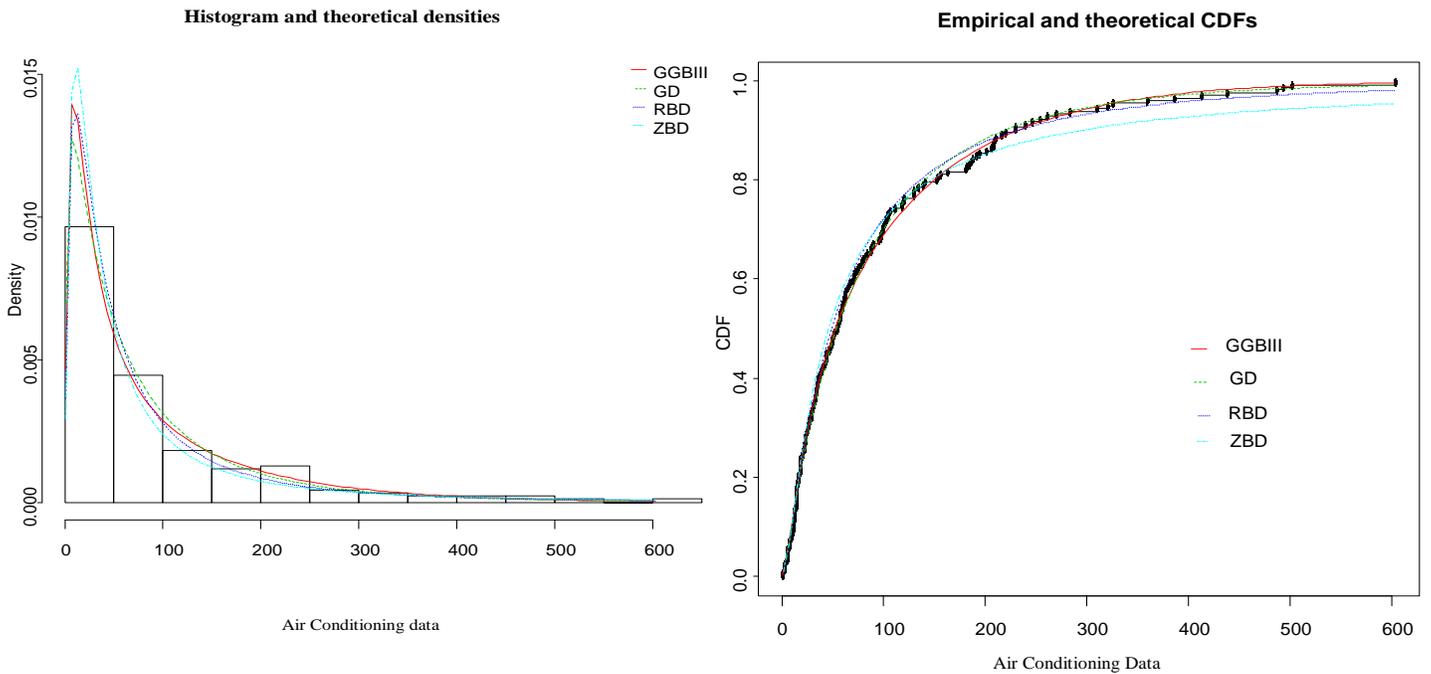

Figure 6: The histogram of the Air conditioning data and the estimated fitted distributions
Figure 7: The cumulative function of the Air conditioning data and the estimated fitted distributions

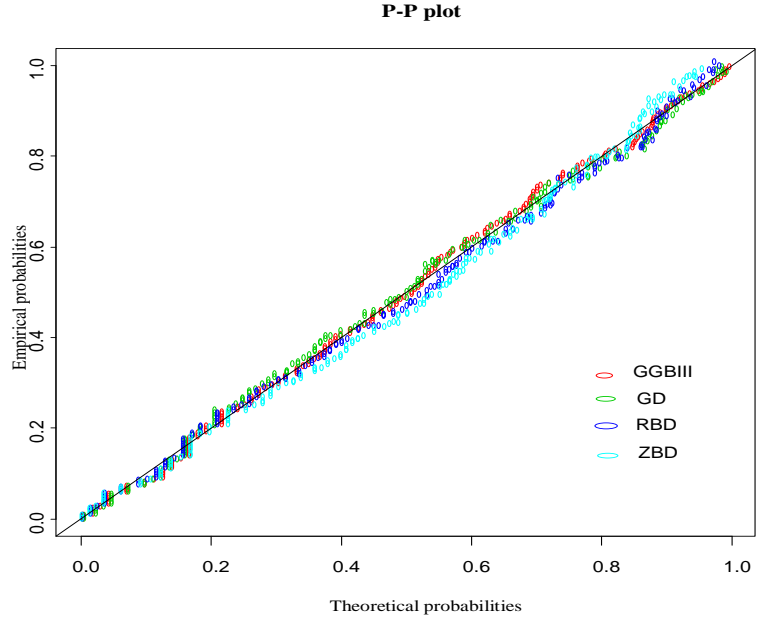

Figure 8: The P-P plot of the Air conditioning data and the estimated fitted distributions

### 6.3 Failure Times of 50 Components (per 1000 hours)

We used a real data set to show that the GGBIII distribution is a better model when compared to GD, RBD, and ZBD distribution. The data set taken from (Murthy et al., 2004) represents the failure times of 50 components (per 1000h):

**Table 10: The Failure Time of 50 Components Data**

```
0.036, 0.058, 0.061, 0.074, 0.078, 0.086, 0.102, 0.103, 0.114, 0.116,
0.148, 0.183, 0.192, 0.254, 0.262, 0.379, 0.381, 0.538, 0.570, 0.574,
0.590, 0.618, 0.645, 0.961, 1.228, 1.600, 2.006, 2.054, 2.804, 3.058,
3.076, 3.147, 3.625, 3.704, 3.931, 4.073, 4.393, 4.534, 4.893,
6.274, 6.816, 7.896, 7.904,  8.022, 9.337, 10.940, 11.020, 13.880,
14.730, 15.080.
```

The asymptotic covariance matrix of the MLEs of the GGBIII model parameters, which is the inverse of the observed Fisher information matrix $I_n^{-1}(\widehat{\mathbf{\Delta}})$ is given by

$$= \begin{pmatrix} 1.028647e-01 & 8.474526e-03 & 0.0140114350 & 9.055202e-04 & 12.591021456 \\ 8.474526e-03 & 3.678572e-03 & 0.0051354478 & 8.182118e-05 & 1.678922408 \\ 1.401144e-02 & 5.135448e-03 & 0.0049282620 & 1.009937e-04 & 2.731526626 \\ 9.055202e-04 & 8.182118e-05 & 0.0001009937 & 9.044504e-08 & 0.001921215 \\ 1.259102e+01 & 1.678922e+00 & 2.7315266258 & 1.921215e-03 & 43.107069967 \end{pmatrix}$$

**Table 11: The Maximum Likelihood Estimation of the Parameters for Failure Time of Components Per Hour**

| MODEL | $\alpha$ | $\beta$ | $\lambda$ | $\delta$ | $p$ | $\theta$ |
|---|---|---|---|---|---|---|
| **GGBIII** | 0.089 (0.067) | 4.079 (2.932) | 0.358 (0.752) | 0.563 (0.179) | 5.618 (3.076) | – |
| **GD** | 0.560 (0.852) | 3.838 (3.722) | 1.681 (1.299) | 0.417 (0.236) | – | 0.104 (0.198) |
| **RB-D** | 6.170 (7.756) | 2.774 (4.290) | 8.018 (12.910) | 0.562 (0.483) | – | – |
| **ZB-D** | 0.076 (0.011) | 5.424 (0.004) | 7.930 (0.019) | 1.331 (0.004) | – | – |

**Table 12: The $-2L$, AIC and BIC of the Distributions for Failure Time of Components per Hour**

| MODEL | $-2Log-Likelihood$ | AIC | BIC |
|---|---|---|---|
| **GGBIII** | 198.4 | 208.4 | 217.9 |
| **GD** | 205.9 | 215.9 | 255.4 |
| **ZB-D** | 207.9 | 215.9 | 223.6 |
| **RBD** | 205.5 | 213.5 | 221.2 |

**Table 13: The Likelihood Ratio Test Statistic**

| MODEL | Hypothesis | LR Statistic |
|---|---|---|
| **GGBIII vs GD** | $H_0: GD \ vs \ H_1: GGBIII$ | 7.5 |
| **GGBIII vs RBD** | $H_0: GD \ vs \ H_1: GGBIII$ | 7.1 |
| **GGBIII vs ZB-D** | $H_0: GD \ vs \ H_1: GGBIII$ | 9.5 |

**Table 14: Goodness of Fit Statistic for Components Data**

| MODEL | Cramer–Von Mises Statistic | Anderson–darling Statistic | Kolmogorov Smirnov Statistic |
|---|---|---|---|
| **GGBIII** | 0.12097 | 0.75666 | 0.10417 |
| **GD** | 0.16636 | 1.02411 | 0.12379 |
| **RB-D** | 0.17121 | 1.05099 | 0.12923 |
| **ZB-D** | 0.14148 | 0.92404 | 0.11922 |

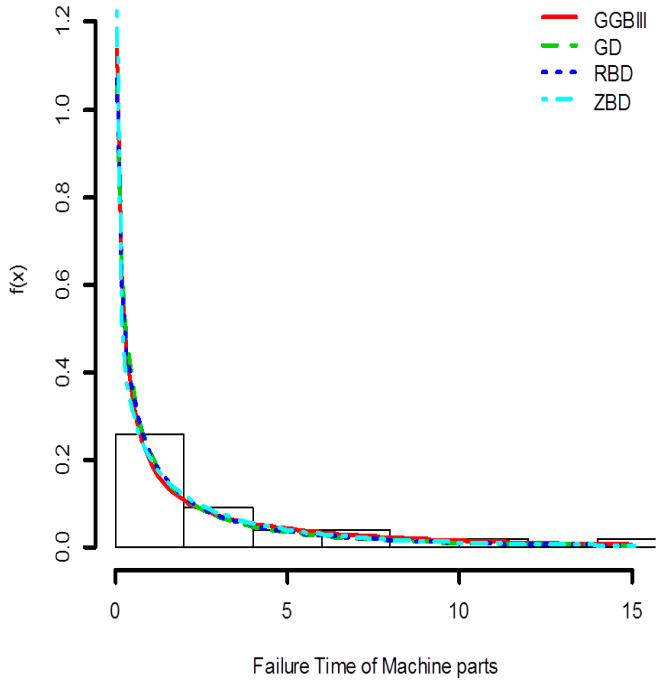
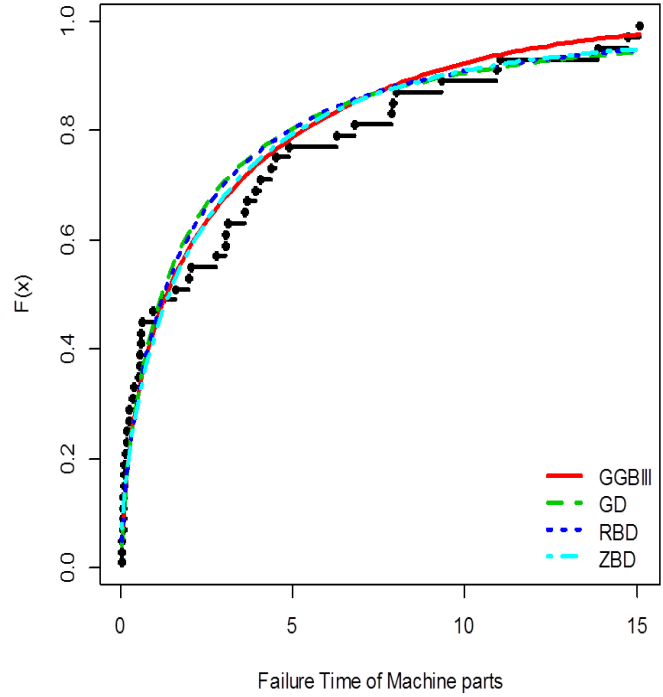
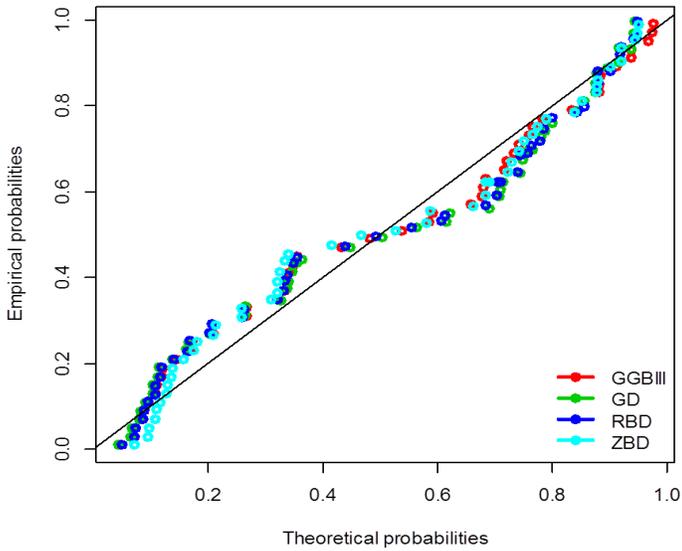
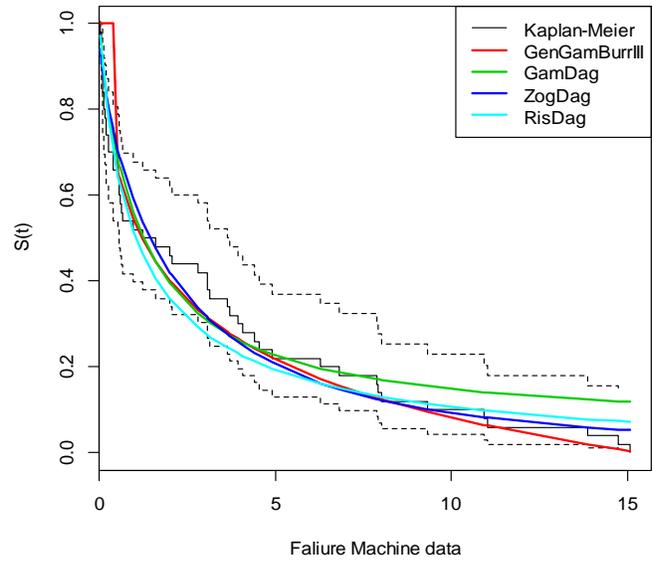

Figure 9: The histogram of the Air conditioning data and the estimated fitted distributions
Figure 10: The Ecf of the Air conditioning data and the estimated fitted distributions
Figure 11: The P-P plot of the Air conditioning data and the estimated fitted distributions
Figure 12: The Esf of the Air conditioning data and the estimated fitted distributions

## 6.4 Breast cancer survival data

We develop an application of the GGBII distribution to a breast cancer data. The study cohort comprises 1207 patients with cancer treated by mastectomy. Patient data were obtained from the database of SPSS software. The data consist of number of months after mastectomy. Uncensored observations correspond to patients having death time computed. Censored observations correspond to patients who were not observed to have died at the time the data were collected. The numbers of censored and uncensored observations are 1135 and 72, respectively, of the total of 1207 patients.

The asymptotic covariance matrix of the MLEs of the GGBIII model parameters, which is the inverse of the observed Fisher information matrix $I_n^{-1}(\hat{\Delta})$ is given by:

$$= \begin{pmatrix} 0.027479129 & 0.0070660643 & -1.08292e-03 & 1.209875e-03 & 0.0932442544 \\ 0.007066064 & 0.0017795919 & -2.87801e-04 & 2.736553e-04 & 0.0193527713 \\ -0.0010829 & -0.000287801 & 3.732178e-05 & -1.02690e-04 & 0.1099804175 \\ 0.001209875 & 0.0002736553 & -1.02690e-04 & 1.710644e-05 & 0.0008351013 \\ 0.093244254 & 0.0193527713 & 1.099804e-01 & 8.351013e-04 & 0.0354401755 \end{pmatrix}$$

**Table 15: The Maximum Likelihood Estimation of the Parameters for Time-to-Death of the Breast Cancer Patients**

| MODEL | $\alpha$ | $\beta$ | $\lambda$ | $\delta$ | $p$ | $\theta$ |
|---|---|---|---|---|---|---|
| **GGBIII** | 0.2523 (0.0363) | 10.1887 (2.3334) | 0.8490 (0.3655) | 0.6490 (0.0456) | 6.0770 (0.5019) | — |
| **GD** | 0.4417 (0.0363) | 32.1491 (2.1059) | 1.3056 (0.0925) | 0.5710 (0.0233) | — | 0.030 (0.0097) |
| **ZB-D** | 0.17549 (0.0052) | 12.0983 (0.0097) | 80.292 (0.01515) | 1.4590 (0.0045) | | |
| **RBD** | 14.3930 (11.7487) | 5.5995 (3.5564) | 370.23 (190.1162) | 0.9590 (0.2591) | | |

**Table 16: THE $-2L$, AIC, AICC AND BIC OF THE MODELS**

| MODEL | $-2Log-Likelihood$ | $AIC$ | $BIC$ |
|---|---|---|---|
| **GGBIII** | 11332.4 | 11342.4 | 11367.9 |
| **GD** | 11439.7 | 11449.7 | 11475.2 |
| **ZB-D** | 11640.8 | 11650.8 | 11671.2 |
| **RBD** | 11454.4 | 11464.4 | 11484.8 |

Table 17: The Likelihood Ratio Test Statistic

| MODEL | Hypothesis | LR Statistic |
|---|---|---|
| **GGBIII vs. GD** | $H_0: GD \ vs \ H_1: GGBIII$ | 107.33 |
| **GGBIII vs. RBD** | $H_0: GD \ vs \ H_1: GGBIII$ | 122.01 |
| **GGBIII vs. ZB-D** | $H_0: GD \ vs \ H_1: GGBIII$ | 308.44 |

Table 18: Goodness of Fit Statistic for Breast Cancer Data

| MODEL | Cramer–Von Mises Statistic | Anderson–darling Statistic | Kolmogorov Smirnov Statistic |
|---|---|---|---|
| **GGBIII** | 0.44366 | 2.18208 | 0.03976 |
| **GD** | 3.05180 | 14.78988 | 0.09135 |
| **RB-D** | 1.86892 | 10.95541 | 0.07256 |
| **ZB-D** | 3.86745 | 23.31508 | 0.10140 |

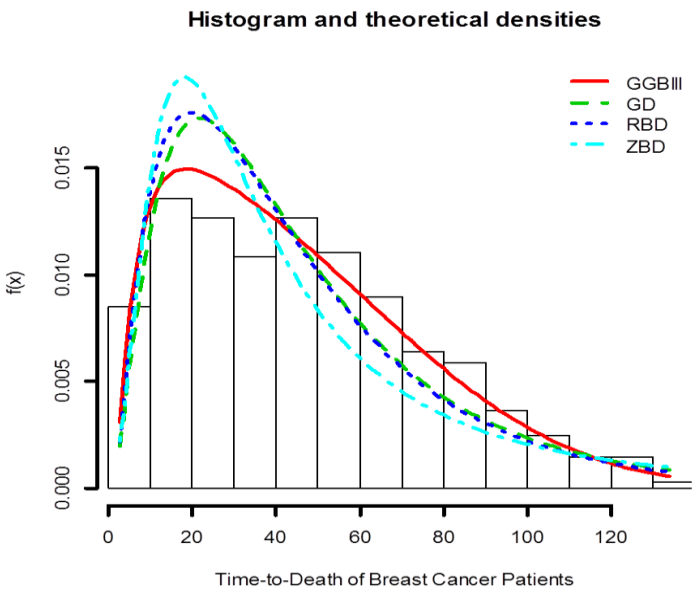
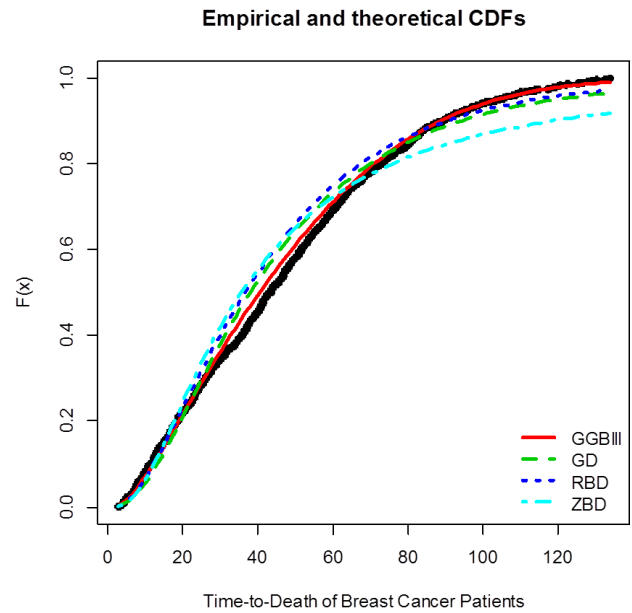

Figure 13: The histogram of the Breast cancer data and the estimated fitted distributions
Figure 14: The Ecf of the Breast cancer data and the estimated fitted distributions

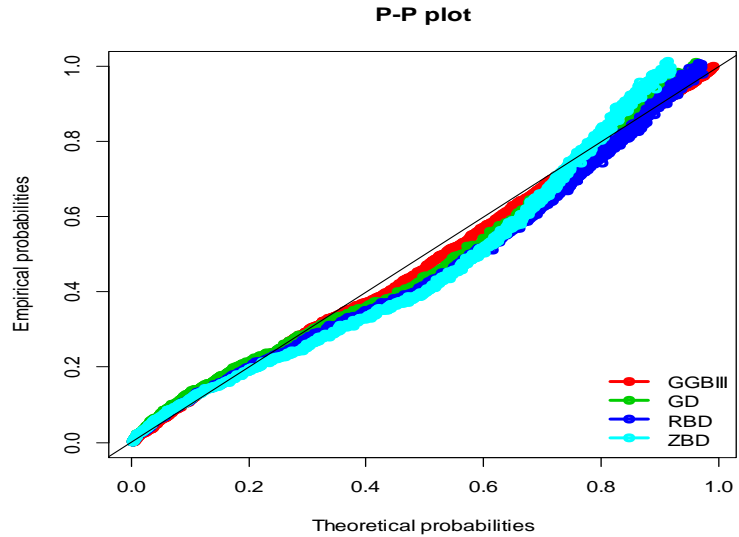

Figure 15: The P-P plot of the Breast cancer data and the estimated fitted distributions

**Summary of the Findings**

A new five-parameter distribution named the Generalized Gamma Burr III distribution has been introduced. It is the generalization of the Burr III distribution.
The proposed distribution has the ability to capture monotonically increasing, decreasing and unimodal hazard rates.
It also reveals that GGBIII distribution has widened the scope of gamma-G family into the area of survival analysis and it has been found amenable in the medical area.
Finally, we showed that the proposed distribution gave the best fit for five well-known data sets (when compared to other distributions including one having five parameters).

## 7. DISCUSSION AND CONCLUSION

A new class of generalized Burr III distribution called the generalized gamma-Burr III distribution is proposed and studied. The idea is to combine two components in a serial system, so that the hazard function is either increasing or more importantly bathtub shaped and unimodal shaped. The GGBIII distribution has the family of Zografos and Balakrishnan distribution as special cases. The density of this new class of distributions was expressed as a linear combination of Burr III density functions. The GGBIII distribution possesses hazard function with flexible behavior. We also obtained closed form expressions for the moments, mean and median deviations, and distribution of order statistics. Maximum likelihood estimation technique was used to estimate the model parameters.
Moreover, to have a strong evidence for this work, the goodness of fit plot for each dataset was provided in order to check how fit the proposed model is to the dataset as compared to other models. Finally, the GGBIII model was applied to FOUR different types of real datasets to illustrate the usefulness and robustness of the distribution in different areas including medical areas and also the model outperformed the existing models of Gamma-Generated Family and is found better than GD and ZBD and RBD which have been fitted to the data used except the breast cancer data. Also, the new model was specifically applied to censored data and it was found more flexible. This paper introduced for the first time the usefulness of the new distribution in survival analysis aside finance, economic, and reliability studies, and it was found that there is a wide significant different in the breast cancer data analysis that is the new distribution is useful in survival analysis far more than other generalizations through Gamma generated family. Thus, we conclude that GENERALIZED GAMMA BURR III is an alternative distribution to gamma families.

**Acknowledgement**